\documentstyle[12pt,epsf]{article}
\title{Supersymmetric derivation of the hard core deuteron's bound state}

\bigskip \bigskip \bigskip

\author{J. Oscar Rosas-Ortiz\\[1.5ex]
       \small {\it Departamento de F\'\i sica Te\'orica}\\
  \small {\it Universidad de Valladolid, 47011 Valladolid, Spain}\\[1ex]
     \small and\\[1ex]    
      \small {\it Departamento de F\'\i sica}, CINVESTAV-IPN,\\
        \small {\it A.P. 14-740, M\'exico 07000 D.F., Mexico} }

\bigskip \bigskip
         
\date{}

\begin{document}

\maketitle

\thispagestyle{empty}
\begin{abstract} 
A supersymmetric construction of potentials describing the hard core
interaction of the neutron-proton system for low energies is proposed. It
considers only the binding energy case and uses the approximation of the
Yukawa potential given by Hulth\'en. Recent experimental data for the
binding energy of the deuteron are used to give the involved orders of
magnitude.
\end{abstract}

\bigskip\bigskip

{\footnotesize
\begin{description}
\item[Key-Words:] Supersymmetry, Deuteron binding energy
\item[PACS:] 34.20.Cf, 03.65.Ge, 03.65.Ca

\end{description}}

\vfill

\hfill 
{\bf Preprint UVA-ES-10-99 }

\newpage
\setcounter{page}{1}

The neutron-proton (n-p) system is the simplest of the composite nuclei. 
It can exist in a stable bound state, the deuteron. This is a nuclide for
which no excited bound states are known, {\it i.e.}, it exists only in its
ground state. For a n-p system in which there are no electrostatic forces
({\it i.e.}, the force due to the two magnetic moments can be considered
as small as an irrelevant correction to the nuclear forces) the
interaction is depicted by a short range central force. Forms of the
related potentials commonly used are the square well, gaussian and Yukawa
potentials (see for example \cite{Pre75}.) Among them, the gaussian and
the square well potentials have deserved special attention because their
mathematical simplicity. On the other hand, the Yukawa potential has a
deeper theoretical significance because its close connection with the
`exchange forces' responsible of the binding between nucleons. 
Historically, this potential is often quoted by the prediction of the
existence of the $\pi$ meson. Hulth\'en noticed that the Yukawa potential
can be conveniently approximated by the expression \cite{Ros48}
\begin{equation} 
\label{hulthen} 
V (r) = - \frac{{\cal V}_0}{e^{r/\alpha} -1},
\end{equation} 
where $r$ is the distance between the nucleons, $\alpha >0$ is a fixed
length (the {\it range} of the potential) and ${\cal V}_0 >0$ is a fixed
energy (the {\it strength} of the potential). 

For low energies, the Schr\"odinger equation for the Yukawa potential is
not solvable analytically and the potential (\ref{hulthen}) is used as the
{\it stand in} for it (see \cite{Lam71} and references quoted therein). In
this range of energies the problem of describing the n-p interactions can
be characterized in two forms: that concerning to bound states or to
scattering states. In the following, we shall only focus on the former
case. On the other hand, empirically is known that the deuteron has a
total spin 1. The obvious interpretation is that the spins of the neutron
and the proton are parallel in the ground state, which would thus be
described by a ${}^3 S_1$-state. Therefore, we will restrict ourselves to
the study of the nuclear binding energy $S$-states. 

Let us stress now that the nucleons cannot approach each other closer than
a certain distance, otherwise, the nucleus would not show an almost
constant nuclear density. Then we must resort to phenomenological
arguments \cite{Hul57}. The usual approach is to modify the nuclear
potential at small distances to be consistent with the experimental data.
A simpler approach considers the so-called {\it hard core model} which is
characterized by a short range infinite repulsion inside the attractive
nuclear interaction \cite{Jas51}.  In other words, a {\it realistic}
potential must contain more than a radial term of the form
(\ref{hulthen}): it must also include an infinitely high barrier term. 

The present paper investigates the supersymmetric nature of the hard core
deuteron's binding energy by doing calculations on the Hulth\'en's
potential. We shall show that the new potential so derived presents a
repulsive barrier term and can be either isospectral (e.g. it shares the
same spectrum) or almost isospectral (the same spectrum except the ground
state) to the Hulth\'en's potential. As a particular case, the new
potential is chosen to be a representation of the hard core nuclear force
describing the n-p low energy binding interactions. The involved orders of
magnitude are given by using the experimental data of $ -2.22456614 (41)
\, {\rm MeV}$ for the deuteron binding energy ${\cal E}_d$ recently
reported in \cite{Kes99}. 

As usual for potentials depending on $r$, the Schr\"odinger equation for
the Hulth\'en's potential (\ref{hulthen}) reduces to an eigenvalue
equation for a particle in a one dimensional effective potential
$V_\ell(x) = \ell(\ell +1)/x^2 - V(x)$, where $\ell$ is the azimuthal
quantum number and $x= r/\alpha$ is a dimensionless radial coordinate. We
are looking for the S states of binding energy ({\it i.e.}, negative
energies and $\ell =0$), therefore the following equation holds
\begin{equation}
\left[ \frac{d^2}{dx^2} + \frac{V_0}{e^x -1} -k^2 \right] \psi(x)  = 0,
\label{schrodinger}
\end{equation}
where $\psi(x) \equiv x R(x)$, with $R(x)$ the standard radial
wavefunction, and $E = - k^2 = (2\mu \alpha^2/ \hbar^2){\cal E}$, $V_0 =
(2\mu \alpha^2 /\hbar^2) {\cal V}_0$, with $\mu$ the reduced mass.

The standard procedure of solution carries out the eigenfunctions
\cite{Flu74}
\begin{equation}
\psi_n(x) = C_n \,\,e^{-k x} (1 - e^{-x}) \, {}_2F_1(2 k + 1 + n, 1 - n, 2
k + 1; e^{-x} ), \quad n=1,2,\dots,
\label{eigenfunction}
\end{equation}
with $C_n$ a normalization constant
\[
C_n \equiv \alpha^{-3/2} \, \frac{\Gamma(n + 2 k)}{\Gamma(n+1) \Gamma(2k
+1)} \, [2k (n + k)(n + 2 k)]^{1/2}.
\]
The corresponding eigenvalues are given by
\begin{equation}
\label{eigenvalue}
E_n = -k^2_n = - \left( \frac{V_0 - n^2}{2n} \right)^2, \quad V_0> n^2,
\quad n=1,2,\dots
\end{equation}
Remark that the problem involves two mutually dependent parameters, $V_0$
and $k$. In practice, the eigenvalue of the energy may be given by
experiment while the strength $V_0$ of the potential is to be determined.
Therefore, equation (\ref{eigenvalue}) can be used to evaluate $V_0$ in
terms of $E_n$. 

The dotted curve on Figure~1 represents the potential (\ref{hulthen}) 
allowing only one bound state, the deuteron's ground state, labeled by a
subscript $H$. In Figure~2 we have plotted the well known corresponding
probability density $\vert \psi_H (x) \vert^2$. 

As regards the supersymmetric scheme, we have in the first place to look
for a superpotential $w(x)$ solving the Riccati equation
\begin{equation}
\label{riccati}
-w'(x) + w^2(x) = V(x) - \epsilon, 
\end{equation}
where the prime denotes derivative with respect to $x$ and the {\it
factorization energy} $\epsilon = -\kappa^2$ is, in principle, any real
number.  By a simple calculation we get for the particular solution
\begin{equation}
\label{particular}
w(x) = \kappa - \frac{1}{e^x -1}, \qquad \kappa > 0,
\end{equation}
which is useful in the cases when the strength can be rewritten as $V_0 =
1 + 2 \kappa$. As usual, the susy partner $\widetilde V$ of $V$ is given
by the {\it shape invariance} condition \cite{Coo95}
\begin{equation}
\label{shape}
\widetilde V (x) = V(x) + 2 w'(x),
\end{equation}
leading to
\begin{equation}
\label{partner}
\widetilde V (x) = -\frac{1+2\kappa}{e^x-1} + \frac{1}{2 \sinh^2 (x/2)}. 
\end{equation}
Potential (\ref{partner}) is a well known result in susy quantum mechanics
(see \cite{Lah88}). It has been used to study susy phase-equivalent
potentials \cite{Tal92} and to establish some interesting connections
between the susy and the variational method \cite{Dri95}.

Observe now the appearance of the r.h.s. term in (\ref{partner}). This
term presents a singularity of order $2/x^2$ at origin and behaves just as
a repulsive centrifugal term with $\ell =1$ in the neighborhood of $x=0$. 
\begin{equation}
\label{origen}
\widetilde V(x) \sim -\frac{1 + 2 \kappa}{x} + \frac{2}{x^2}, \quad x<< 1. 
\end{equation}
As the value $x=1$ implies $r = \alpha$, the approximation (\ref{origen}) 
holds in the range of the initial potential $V(x)$. In the region $x>1$,
the potential $\widetilde V(x)$ rapidly becomes negligible (see Figure~1.)
Here, the strength $V_0 =1 + 2 \kappa$ plays the role of a coupling
constant. 

The above results can be used to determine the eigenfunctions and
eigenvalues connected with the new potential $\widetilde V(x)$. The
procedure consits now in factorizing the related Hamiltonias by
\begin{equation}
\label{factor}
H = A^{\dagger} A + \epsilon, \qquad \widetilde H = A A^{\dagger} + 
\epsilon,
\end{equation}
with
\begin{equation}
\label{operator}
A \equiv \frac{d}{dx} + w(x).
\end{equation}
It is straightforward to check that equations (\ref{riccati}),
(\ref{shape}) and (\ref{operator}) automatically lead to (\ref{factor}).
Then, an intertwining relationship holds:
\begin{equation}
\label{intertwining}
\widetilde H A = A H.
\end{equation}
If $\psi(x)$ is an eigenfunction of $H$ with eigenvalue $E$, equation
(\ref{intertwining}) gives
\[
\widetilde H (A \psi) = E (A \psi), \qquad A \psi \neq 0.
\]
Therefore, if $\psi \in L^2({\bf R})$, we get the normalized eigenstate of
$\widetilde H$
\begin{equation}
\label{susyfunction}
\widetilde \psi(x) = (E - \epsilon)^{-1/2} A \psi(x).
\end{equation}
Now, let us stress on the information displayed by equations
(\ref{factor})--(\ref{susyfunction}). First, in the case when $\epsilon =
E_n$, for any $n=1,2,\dots$, the l.h.s. equation in (\ref{factor}) applies
on $\psi_n(x)$ as $H \psi_n(x) = E_n \psi_n(x)$, and consequently
$A^{\dagger} A \psi_n(x)=0$. It is easy to check that $A \psi_n(x)=0$ is a
sufficient condition to get square integrable functions. Therefore,
equation (\ref{susyfunction}) means that $\psi_n(x)$ has not a susy
partner $\widetilde \psi_n(x)$ and the couple of Hamiltonians
(\ref{factor}) corresponds to a case of unbroken supersymmetry
\cite{Coo95}.

On the other hand, when $\epsilon \neq E_n$ for every $n=1,2,\dots$, there
is no square integrable eigenfunction of $H$ annihilated by $A$, and
equation (\ref{susyfunction}) means that every $\psi_n(x)$ will have a
susy partner $\widetilde \psi_n(x)$. Now, from the r.h.s. of
(\ref{factor}), it is clear that a function $\widetilde \psi_*(x)$,
obeying $A^{\dagger} \widetilde \psi_* = 0$, leads to $\widetilde H
\widetilde \psi_*(x) = \epsilon \widetilde \psi_*(x)$. Hence, if
$\widetilde \psi_* \in L^2({\bf R})$, it must be added to the new set $\{
\widetilde \psi \}$. When $\widetilde \psi_*(x)$ is an square integrable
function the Hamiltonians (\ref{factor}) present unbroken supersymmetry,
otherwise they correspond to a case of broken supersymmetry \cite{Coo95}.

For the superpotential we are dealing with, one gets $\widetilde \psi_*(x) 
\propto e^{\kappa x} (1 - e^{-x})^{-1}$, which is obviously not square
integrable in $[0, \infty)$ for $\kappa >0$. Then, the supersymmetric
behaviour of the Hamiltonians (\ref{factor}) lies in the selection of
$\epsilon$, {\it i.e.}, if $\epsilon$ is chosen to be either an eigenvalue
of $H$ or not. 

In the following we shall consider the case when the initial potential
$V(x)$ allows the binding of only two states. The purpose of this
convention will be apparent in the sequel. To get a system with only two
energy levels, the strength of the Hulthen's potential has to be in the
domain $4< V_0 < 9$. Providing these values of $V_0$ we get for the
energies: $-16 < E_1 < -2.25$, and $-1.56 < E_2 <0$. 

In order to get an idea of the orders of magnitude involved, we note that,
although there is no {\it a priori} reason why $\alpha$ should not be
different for different sorts of the stationary systems described by
$V(x)$, we can take its numerical value as $\alpha =3 \mathaccent23f$. 
Such assertion is justified by the fact that the mean distance between
nucleons ({\it i.e.}, the {\it size} of the nucleus) is in the range of 2
or 4 $\mathaccent23f$ \cite{Pre75}. Therefore, we get $(\hbar^2 /\alpha^2
m_p) \simeq 4.6113 \, {\rm MeV}$.  Here, we have assumed that the neutron
and proton masses are equal to $m_p$, hence $2 \mu = m_p$. In this way,
the experimental value of the deuteron binding energy ${\cal E}_d$ becomes
in a dimensionless value $E_d \simeq -0.4825$ ($k_d \simeq 0.6946$).
Remark that, for the above selected domain of $V_0$, the deuteron energy
lies in the domain of the exited state $E_2$ and not in the domain of the
ground state $E_1$. 

Let us consider now the factorization energy fixed as $\epsilon = E_1 =
-k_1^2$. In this case, as discussed above, we will have a couple
(\ref{factor}) with unbroken supersymmetry. Hence $A \psi_1(x) = 0$, and
the potential $\widetilde V (x)$, with $V_0 \equiv 1 + 2 \kappa = 1 + 2
k_1$, misses the ground state of $V(x)$ and admits only one bound state
(see equation (\ref{susyfunction})): 
\begin{equation}
\label{binding}
\widetilde \psi (x) \equiv (E_2 - E_1)^{-1/2} \left[ \frac{d}{dx} \ln
\psi_2(x) + w(x) \right] \psi_2(x),
\end{equation}
with eigenvalue
\begin{equation}
\label{susyenergy}
E_2 = - \left( \frac{V_0 -4}{4} \right)^2 = - \left(
\frac{2 k_1 -3}{4} \right)^2.
\end{equation}

We go a steep further and impose that the numerical value of $E_2$ be
determined by experiment and let it be equal to $E_d$, the dimensionless
value of the deuteron binding energy, therefore
\begin{equation}
\label{numerical}
k_1 = \frac{4 k_d +3}{2} \simeq 2.8892, \qquad V_0 = 4 k_d +4 \simeq
6.7784
\end{equation}
which agrees with the previously stated domains for $V_0$, $E_1$ and
$E_2$. Then, an initial potential (\ref{hulthen}), with range $\alpha = 3
\mathaccent23f$ and strength ${\cal V}_0 \simeq 31.2572 \, {\rm MeV}$, has
a susy partner (\ref{partner}) allowing the binding of a single state
$\widetilde \psi (x)$ with an energy exactly equal to ${\cal E}_d$. The
main characteristic of this new potential is its centrifugal term, which
reduces the range of approaching between the nucleons. Therefore, we have
constructed a radial potential describing the hard core binding
interaction between the nucleons.

The probability density connected with the ground state (and single!) 
eigenfunction (\ref{binding}) of the hard core Hamiltonian $\widetilde H$,
for the numerical data mentioned above, has been plotted on Figure~2. Its
features can be contrasted with those of the {\it no core} Hamiltonian
$H$, labeled by $\vert \psi_H(x) \vert^2$, in the same figure. Observe the
displacement to the right of $\widetilde \psi$ with respect to $\psi_H$.
An easy calculation shows that $\widetilde \psi (x) \sim (1 - e^{-x}) \,
\psi_H (x)$, hence, near the origin, $\widetilde \psi$ goes to zero as
$x^2$ whereas $\psi_H$ goes just as $x$, and the probability to find the
state $\widetilde \psi(x)$ near to zero is minor than the probability to
find $\psi_H(x)$ at the same place.

We shall now discuss some of the various implications of our results. 
First, the susy procedure accomplishes the derivation of a short range
potential $\widetilde V(x)$ allowing only one bound state. This potential
exhibits many of the qualitative features concerned with a hard core
potential. It behaves predominantly as an effective radial potential
$V_{\ell}(x)$, with $\ell =1$, near the origin (see equation
(\ref{origen})) and goes negligible with the increasing of $x$ in the
region $x>1$. 

Second, from Figure~2, it is clear that there is a considerable
probability of finding the two nucleons at distances larger than $\alpha$.
Therefore, the nuclear force connected with $\widetilde V(x)$ plays a
relevant role only in a weak way because the neutron and the proton are
outside each other's range so much of the time. That is, of course, a well
known feature in the behaviour of the deuteron's ground state
\cite{Pre75,Lev90}. It is interesting to remark on the smoothest of
$\widetilde \psi(x)$, this is a nodeless function, just as one might
expect for the eigenfunction of a single stable bound state. Therefore,
the wavefunction (\ref{binding}) corresponds to the hard core ground state
eigenfunction of the deuteron. 

In summary, the method developed in the paper relied on the assumption
that it is possible to describe the interacting system of a proton plus a
neutron by a Schr\"odinger equation. The method allowed the construction
of a hard core potential for the deuteron and the estimation of the
potential energy necessary to give the observed deuteron's binding energy. 
It is remarkable that even very refined experiments at low energies do not
suffice to determine more than the range and strength of the involved
potential, leading the detailed shape completely indeterminate.

Observe that the hard core hypothesis makes the strength
(\ref{numerical}), of the potential (\ref{hulthen}), increased by $V_0
\simeq 3 V_{0H}$, where $V_{0H} = 2 k_d + 1$. In general, the nuclear
forces are quite complicated and the assumption of a pure ${}^3S$-state
does not suffice to explain the deuteron quadrupole moment, this makes
necessary to introduce a tensor force \cite{Hul57}. However, it is well
known that the Hulth\'en's potential gives a good approximation for the
binding energy in the terms discussed at the very begining of the paper. 
On the other hand, as mentioned above, there are different ways to modify
the nuclear potential in order to get a hard core model, the potential
derived here could be tested by the nucleon-nucleon scattering approaches,
where the hard core hypothesis seems to be compatible with the empirical
data \cite{Jas51}. 

As regards the unbroken susy results, we have to say that recent susy
treatments, involving general solutions to the Riccati equation
(\ref{riccati}), have shown the way to introduce susy partner potentials
sharing exactly the same spectra \cite{Mie84}. That is, we could obtain a
new Hamiltonian $\widetilde H$ which, joined in a susy couple with $H$,
present unbroken supersymmetry by extending the superpotential
(\ref{particular}) to a general solution of (\ref{riccati}). Although that
method has been sucssesfully applied in the study of some interesting
potentials (see e.g.  \cite{Fer98,Ros98}), it is out of the present scope
and will be given elsewhere. 

I would like express my gratitude to L.M.~Nieto for his unending patience
during the long discussions concerning this paper. I had learnt quite a
lot of him and J.~Negro during my scientific stay in Valladolid (Spain). I
am also greatly indebted to Miss M. Lomeli for her constant help in typing
the manuscript. 

This work has been partially supported by Junta de Castilla y Le\'on,
Spain (project C02/97). The kind hospitality at Departamento de
F\'{\i}sica Te\'orica, Universidad de Valladolid, Spain, is also
aknowledged. 


\newpage

\begin{figure}[htbp]
\centerline{
\epsfxsize=14cm
\epsfbox{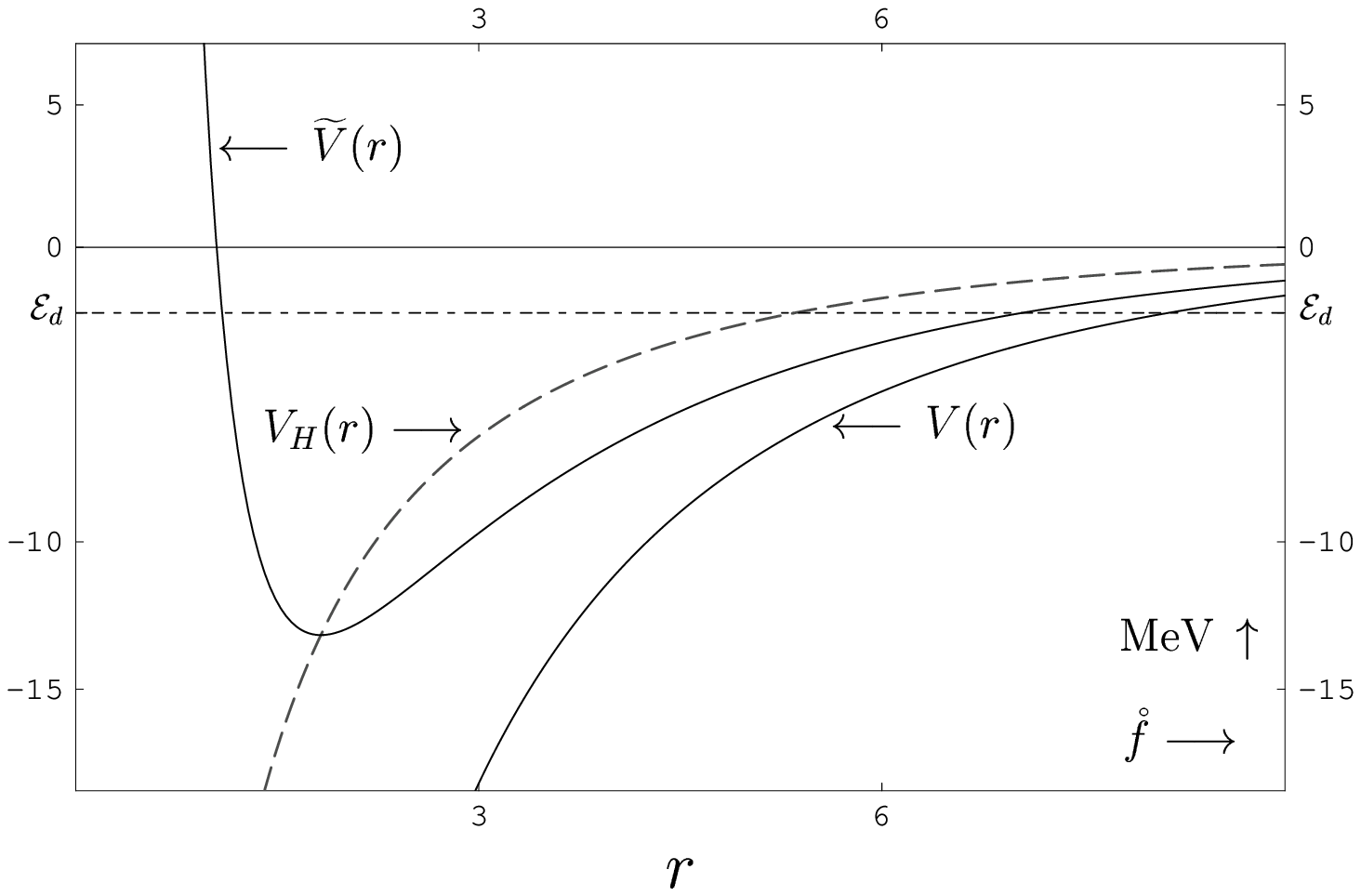}}
\begin{center}
\begin{minipage}{12cm}
\caption{
The susy partner potentials (\ref{hulthen}) and (\ref{partner}) with
$\alpha = 3 \mathaccent23 f$ and ${\cal V}_0 \simeq 31.2572 \, {\rm MeV}$.
The deuteron's binding energy ${\cal E}_d \simeq -2.2245 \, {\rm MeV}$ has
been ticked on the frame. The potential $V_H$ has a strength ${\cal V}_H
\simeq 11.0173 \, {\rm MeV}$ and the same value of $\alpha$. 
}
\end{minipage}
\end{center}
\end{figure}

\newpage

\begin{figure}[htbp]
\centerline{
\epsfxsize=14truecm
\epsfbox{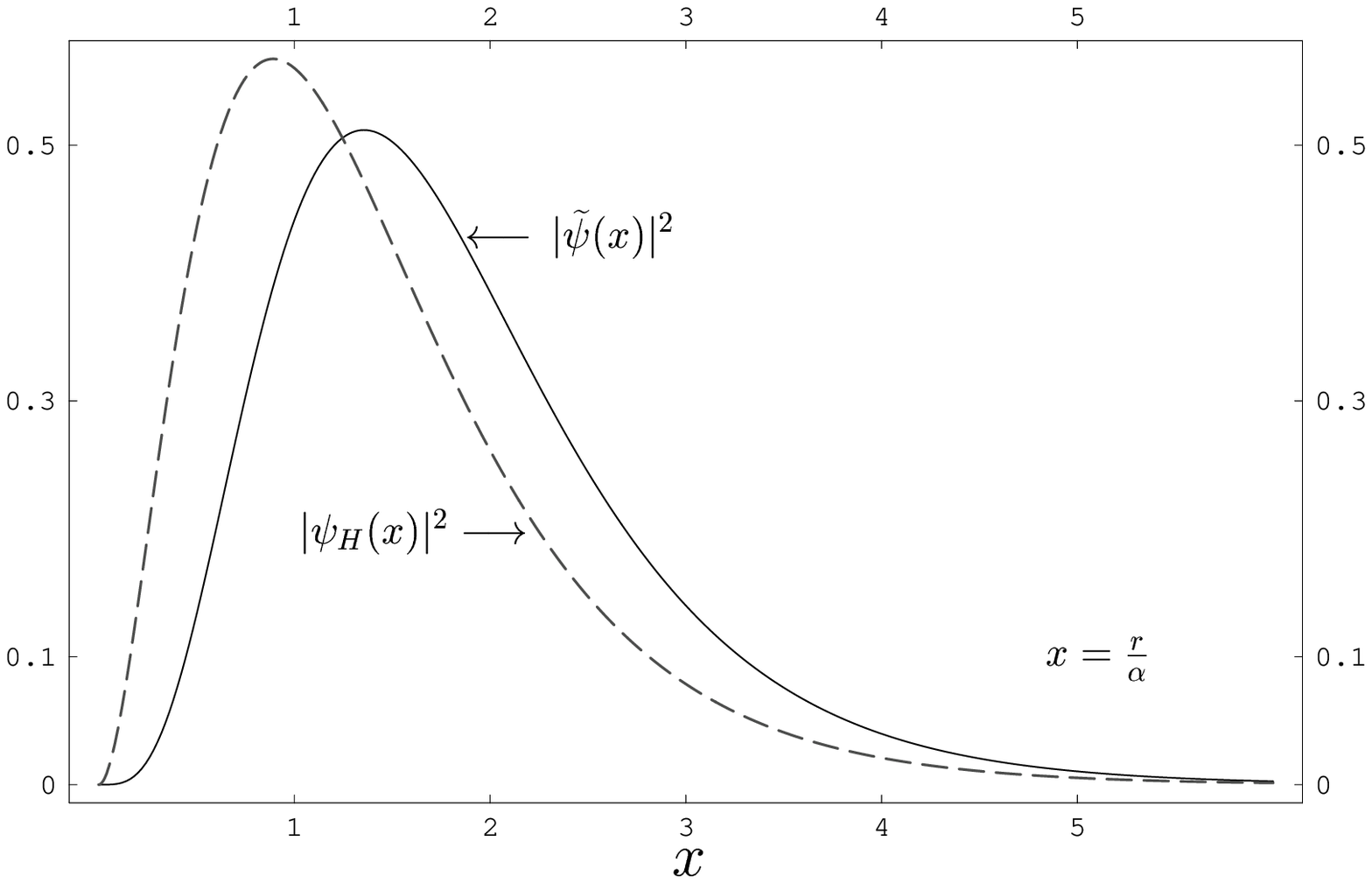}}
\begin{center}
\begin{minipage}{12cm}
\caption{
The deuteron's ground state probability density for the cases without core
$\vert \psi_H(x) \vert^2$, and hard core $\vert \widetilde \psi (x) 
\vert^2$, with $\widetilde \psi(x)$ given in (\ref{binding}) and the
values displayed in Figure~1. Observe the displacement to the right of
$\vert \widetilde \psi (x)  \vert^2$ with respect to $\vert \psi_H(x) 
\vert^2$.
}
\end{minipage}
\end{center}
\end{figure}


\begin{thebibliography}{99}

\bibitem{Pre75}
M.A. Preston and R.K. Bhaduri, {\em Structure of the nucleus},
(Addison, Reading, Mass. 1975); J.M. Blatt and V.E. Weisskopf, {\em
Theoretical Nuclear Physics}, (Springer, N.Y. 1979) 

\bibitem{Ros48}
L. Hulth\'en, Phys. Rev. {\bf 61}, 671 (1942); L. Rosenfeld, {\em Nuclear
Forces}, (North-Holland, Amsterdam 1948)

\bibitem{Lam71}
C.S. Lam and Y.P. Varshni, Phys. Rev. A {\bf 4}, 1875 (1971)

\bibitem{Hul57}
L. Hulth\'en and M. Sugawara, {\em The two-nucleon problem} in
Encyclopedia of Physics, edited by S. Fl\"ugge (Springer, Berlin 1957) 
vol. {\bf 39}, p. 1. 

\bibitem{Jas51}
R. Jastrow, Phys. Rev. {\bf 81}, 165 (1951).

\bibitem{Kes99}
E.G. Kessler {\it et. al.}, Phys. Lett. A {\bf 255}, 221 (1999)

\bibitem{Flu74}
S. Fl\"ugge, {\em Practical Quantum Mechanics} (Springer, Berlin 1971); 
O.L. De Lange and R.E. Raab, {\em Operator Methods in Quantun Mechanics}
(Claredon Press, Oxford 1991) 

\bibitem{Coo95}
F. Cooper, A. Khare and U. Sukhatme, Phys. Rep. {\bf 251}, 267 (1995)

\bibitem{Lah88}
U. Laha, C. Bhattacharyya, K. Roy and B. Talukdar, Phys. Rev. C {\bf 38},
558 (1988)

\bibitem{Tal92}
B. Talukdar, U. Das, C. Bhattacharyya and P.K. Bera, J. Phys. A {\bf 25},
4073 (1992).

\bibitem{Dri95}
E. Drigo Filho and R.M. Ricotta, Mod. Phys. Lett. A {\bf 10}, 1613 (1995)

\bibitem{Lev90}
J.M. L\'evi-Leblond and F. Balibar, {\em Quantics} (North-Holland,
Amsterdam 1990)

\bibitem {Mie84}
B. Mielnik, J. Math. Phys. {\bf 25}, 3387 (1984); D.J. Fern\'andez C.,
Lett. Math. Phys. {\bf 8}, 337 (1984); M.M.~Nieto, Phys. Lett. B {\bf
145}, 208 (1984)

\bibitem{Fer98}
D.J. Fern\'andez, L.M.~Glasser and L.M.~Nieto, Phys. Letts A {\bf 240}, 15
(1998); D.J.~Fern\'andez C. and V.~Hussin, J. Phys. A {\bf 32}, 3603
(1999)

\bibitem{Ros98}
J.O. Rosas-Ortiz, J. Phys. A {\bf 31}, L507 (1998); J. Phys. A {\bf 31},
10163 (1998). 
\end{thebibliography}
\end{document}